\documentclass[showpacs,showkeys,preprintnumbers]{revtex4}
\usepackage[ngerman,english]{babel}
\usepackage[latin1]{inputenc}
\usepackage{amsmath,amssymb,amstext,amsfonts}
\usepackage{mathrsfs,color}
\usepackage{dsfont}
\usepackage{bbm}
\usepackage{amsthm}
\usepackage{graphicx}
\usepackage{subfigure}
\usepackage{multirow}
\usepackage{framed}
\usepackage{hyperref}
\usepackage{braket}
\usepackage{simplewick}
\usepackage{feynmp}
\begin{document}

\title{Low-energy limit of the extended Linear Sigma Model}
\author{Florian Divotgey$^{a}$, Peter Kovacs$^{b}$, Francesco Giacosa$^{a,c}$, Dirk H.\
Rischke$^{a}$}
\affiliation{$^{a}$Institut für Theoretische Physik, Johann Wolfgang Goethe--Universität, 
Max-von-Laue-Str.\ 1, 60438 Frankfurt am Main, Germany}
\affiliation{$^{b}$Institute for Particle and Nuclear Physics, Wigner Research
Centre for Physics, 
Hungarian Academy of Sciences, H-1525 Budapest, Hungary}
\affiliation{$^{c}$Institute of Physics, Jan-Kochanowski University, 
ul.\ Swietokrzyska 15, 25-406 Kielce, Poland} 
\begin{abstract}
The extended Linear Sigma Model (eLSM) is an effective hadronic model based on the linear realization
of chiral symmetry $SU(N_f)_L \times SU(N_f)_R$, with (pseudo)scalar and (axial-)vector mesons as 
degrees of freedom. 
In this paper, we study the low-energy limit of the eLSM for $N_f=2$ flavors
by integrating out all fields except for the pions, the (pseudo-)Nambu--Goldstone bosons of chiral
symmetry breaking. We only keep terms entering at tree level and up to fourth order in powers of derivatives of 
the pion fields. Up to this order, there are four low-energy coupling constants in the resulting low-energy effective 
action. We show that the latter is formally identical to 
Chiral Perturbation Theory (ChPT), after choosing a representative for the coset space generated
by chiral symmetry breaking and expanding up to fourth order in powers of derivatives of the pion fields. 
Two of the low-energy coupling constants of the eLSM are uniquely determined by a fit to hadron 
masses and decay widths. We find that their tree-level values are in reasonable agreement with the
corresponding low-energy coupling constants of ChPT. The other two low-energy coupling constants
are functions of parameters that can in principle be determined by $\pi\pi$ scattering, 
which has not yet been studied within the eLSM. Therefore, we use the respective values from 
ChPT to make a prediction for the values of these parameters in the eLSM Lagrangian.
\end{abstract}
\maketitle

\section{Introduction}
\label{intro}

The physics of the strong interaction is described by Quantum
Chromodynamics (QCD). For $N_{f}$ massless quark flavors the classical QCD Lagrangian possesses a global
$U(N_{f})_{L}\times U(N_{f})_{R}\cong U(N_{f})_{V}\times U(N_{f})_{A}$ symmetry.
At quantum level, this symmetry is reduced to $SU(N_{f})_{V}\times SU(N_{f})_{A}\times U(1)_{V}$, 
since the $U(1)_{A}$ symmetry is explicitly broken by a quantum anomaly \cite{tHooft}. 
The $U(1)_{V}$ symmetry corresponds to quark number conservation. Since it is trivially
fulfilled in any theory with hadrons as degrees of freedom, we do not need to consider it
in the following. The remaining $SU(N_{f})_{V}\times SU(N_{f})_{A}$ symmetry, the so-called chiral symmetry,
is explicitly broken to $SU(N_f)_V$ by nonvanishing and equal quark masses,
and to the direct product of $N_f-1$ separate $U(1)$ groups if all quark masses are unequal. 
It is well known that the experimentally
observed hadrons can be grouped into irreducible representations of
$SU(N_{f})_{V}$ and not into those of $SU(N_{f})_{V}\times SU(N_{f})_{A}$
\cite{pdg}. This observation provides strong evidence for the fact that chiral
symmetry must be spontaneously broken to its diagonal flavor subgroup
$SU(N_{f})_{V}$. As a consequence of the spontaneous breakdown of chiral
symmetry, we expect the occurrence of $N_{f}^{2}-1$ (pseudo-)Nambu--Goldstone
bosons. Throughout this work, we restrict ourselves to the two-flavor case, $N_{f}=2$. 
Then, the three (pseudo-)Nambu--Goldstone bosons are
given by the pion isotriplet $\vec{\pi}$.

Another important property of QCD is that its running coupling constant $\alpha_{S}$
becomes large at small energies. This phenomenon implies that nonperturbative
methods are needed to investigate the low-energy spectrum of QCD. Besides
lattice methods, one can also use Effective Field Theories (EFTs) to
investigate the low-energy dynamics of QCD. The most prominent, systematic, and
well-defined approach of this type is Chiral Perturbation Theory
(ChPT), see e.g.\ Refs.\ \cite{gale,leutwyler,scherer,pich,manohar} and refs.\
therein. ChPT is a theory which describes the dynamics of the (pseudo)Nambu--Goldstone bosons, 
i.e., for $N_f=2$ the pions. In ChPT, chiral symmetry is nonlinearly realized, i.e., the Nambu-Goldstone bosons 
enter as parameters of the representative of the coset space $SU(N_f) \times SU(N_f)/ SU(N_f)$ 
of chiral symmetry breaking. ChPT is defined by a Lagrangian containing
all chiral invariants constructed from powers of derivatives of the coset representative. 
The coupling constants multiplying these invariants are the so-called low-energy constants (LECs).
Since the Lagrangian contains an arbitrary number of derivatives of the coset representative,
it is not perturbatively renormalizable. However, a power series in derivatives of the coset representative 
is equivalent to a power series in $p/(4 \pi f_\pi)$, where $p$ is the momentum of the pion field and
$f_\pi$ the pion decay constant. Thus, for small pion momenta this power series is expected
to converge. Moreover, one can remove all infinities order by order in the pion momentum by absorbing 
them into the LECs.
The fundamentals of ChPT were investigated in Ref.\ \cite{leutwyler} where it was
shown that ChPT has the very same Green functions as QCD in the low-energy
limit. In conclusion, ChPT is definitely the best approach that we have to
study the interactions of (slow) pions.

ChPT was extended by including vector mesons, see
e.g.\ Refs.\ \cite{egpr,chptvm}. In the seminal work of Ref.\ \cite{egpr} it was shown
they play an important role in determining the values of the LECs. Yet, ChPT
becomes less and less accurate when the energy scale increases. In
particular, both the scalar (up to 1.7 GeV) and the axial-vector (up to 1.5
GeV) sectors are notoriously problematic due to the existence of broad
resonances [such as $f_{0}(500),$ $f_{0}(1370),$ $a_{1}(1230)$] and of
resonances close to thresholds in (pseudo-)Nambu--Goldstone boson scattering processes  
[i.e., $a_{0}(980)$ and $f_{0}(980)$]. The
scalar sector is also important since it is related to both the chiral and the
gluon condensate (i.e., the vacuum expectation values of the 
chiral partner of the pion and of the scalar dilaton/glueball field).

An alternative approach to the low-energy dynamics of QCD
is given by hadronic models based on a linear realization of chiral symmetry. Such models are usually
referred to as Linear Sigma Models (LSMs), which historically were studied
even before ChPT \cite{lee} [see also Refs.\ \cite{koch,pisarski,geffen}]. 
A significant difference between the two approaches is that in LSMs the chiral partners of 
the (pseudo-)Nambu--Goldstone
bosons appear on an equal footing, i.e., for $N_f =2$ the scalar sigma field $\sigma_N$ enters besides the pseudoscalar
pions. However, the simple LSM with just pions and sigma does not have the same low-energy limit as QCD, i.e.,
its LECs do not assume the same values as in ChPT \cite{gale}. The LSM was extended by (axial-)vector degrees
of freedom in Refs.\ \cite{ko,urban}. More recently, the so-called extended Linear Sigma Model (eLSM) was developed,
which contains all quark-antiquark mesons with (pseudo)scalar and 
(axial-)vector quantum numbers below 2 GeV in mass. 
The Lagrangian of the eLSM is constructed to respect the chiral and the dilatation symmetry of QCD and
to reflect the pattern of their respective breaking in nature.
Requiring dilatation symmetry and demanding that
only positive semi-definite powers of the dilaton field enter the Lagrangian of the eLSM implies that the latter contains
only a finite number of chiral invariants. The eLSM was first presented for $N_{f}=2$
in Refs.\ \cite{denisnf2,staniold} and then enlarged to $N_{f}=3$ in Refs.\
\cite{dick,staninew}. It was also studied for $N_{f}=4$
\cite{walaa} and for baryons in the vacuum \cite{gallas} and at nonzero
density \cite{heinz}. 

A fit of the parameters of the eLSM to experimentally
measured masses and decay widths shows an agreement on the 5\% level \cite{dick}. This is remarkable,
given the simplicity of the assumptions underlying the eLSM.
A natural question then arises: does the eLSM have the same low-energy limit as QCD, i.e.,
does it reproduce the LECs of ChPT? In order to answer this question, we proceed as follows.
On the one hand we choose a definite representation for the coset representative and expand 
the ChPT Lagrangian up to fourth order in powers of derivatives of the pion field. 
Then, the coupling constants of the resulting Lagrangian are well-defined functions of the LECs of ChPT.
On the other hand, we successively integrate out all fields of the eLSM except for the pions. In this paper,
we work at tree-level, i.e., we neglect all loop corrections, and keep terms up to fourth order in the pion fields. 
This enables us to perform this integration
in a completely analytical way. The resulting low-energy
effective action has the same mathematical form as the above mentioned Lagrangian 
resulting from expanding ChPT in powers of derivatives
of the pion field. However, the respective coupling constants are now well-defined functions of the
parameters of the eLSM, most of which were previously determined by the fit of Ref.\ \cite{dick}. 
The question whether the eLSM has the same low-energy limit as QCD thus boils down to how well 
its low-energy coupling constants compare to the values obtained from ChPT.
A positive answer would validate the eLSM as a low-energy model for QCD.

The paper is organized as follows: Sec.\ \ref{chpt} includes a short summary of
$N_f=2$ ChPT. In Sec.\ \ref{elsm} we briefly present the $N_f=2$ version of the
eLSM and show in detail how to derive its low-energy limit. In Sec.\ \ref{res}
we compare the numerical results for the low-energy coupling constants in ChPT and the eLSM.
Finally, in Sec.\ \ref{con} we present our conclusions and an outlook for future studies.
We defer lengthy formulas to App.\ \ref{appa}. 
Appendix \ref{appb} contains a discussion of other scenarios: the case without (axial-)vector
mesons and the case in which the resonance $f_{0}(500)$, a putative four-quark state, is regarded as the
chiral partner of the pion.

\section{Chiral Perturbation Theory}

\label{chpt}

ChPT is a well-defined low-energy EFT of QCD.
It relies on a systematic low-energy analysis of the hadronic
$n$-point functions built from scalar, pseudoscalar, vector, and axial-vector
quark bilinears. The structure of these $n$-point functions is determined
by chiral symmetry, since they have to transform in some representation of
$SU(N_{f})_{V}\times SU(N_{f})_{A}$. In addition to that, chiral symmetry
gives rise to symmetry relations among these $n$-point functions, the
so-called Ward-Fradkin-Takahashi (WFT) identities. These symmetry relations
allow for a systematic analysis of the hadronic $n$-point functions. 

Another important property of the hadronic $n$-point functions is that they
always have a pole whenever an intermediate particle can be created on-shell.
In the case of QCD, the pole with the smallest energy that one observes corresponds to an on-shell pion,
which shows that the low-energy dynamics of QCD is determined by the
interactions of the pions among themselves. The idea behind ChPT is to perform
a so-called chiral expansion, i.e., a simultaneous expansion in powers of
quark masses and pion momenta of the QCD generating functional (with 
external fields coupling to the above mentioned quark bilinears)
\begin{equation}
Z_{QCD}=Z_{2}+Z_{4}+\ldots\text{ ,} \label{chpt1}%
\end{equation}
where the different terms in this expansion $Z_{2n}$, $n=1,2,\ldots$, include all possible
combinations of the coset representative of chiral symmetry breaking and the
external fields which are allowed by local chiral symmetry as well as by
CPT and proper orthochronous Lorentz transformations. It is therefore clear
that the number of allowed interaction terms rapidly increases with the order of
the expansion. Up to and including next-to-leading order (NLO), the most general chiral Lagrangian is
given by
\begin{equation}
\mathscr{L}_{\chi PT}=\mathscr{L}_{2}+\mathscr{L}_{4}\text{ ,} \label{chpt2}
\end{equation}
where the leading-order (LO) and NLO terms of the Lagrangian are given by Eqs.\ (\ref{app1})
and (\ref{app2}) in App.\ \ref{appa1}. At LO the chiral Lagrangian contains
only two free parameters, the pion decay constant $f_{\pi}$ and the constant
$B_0$, which is related to the bare quark mass. At NLO the number of free
parameters increases to ten. In this case, one has seven LECs $\ell_{i}$, $i=1,\ldots,7$, 
and three additional coupling constants
$h_{i}$, $i=1,2,3$, see Eq.\ (\ref{app2}).

In this work, we are interested in the detailed interaction structure of the
pion fields among themselves. Therefore, we choose
\begin{equation}
\mathcal{U}=\frac{1}{f_{\pi}}\left(  \sigma+i\pi_{i}\tau^{i}\right)  \text{
with }\sigma=f_{\pi}\sqrt{1-\pi_{i}^{2}/f_{\pi}^{2}} \label{chpt3}%
\end{equation}
as a parametrization of the coset space $SU(2)\times SU(2)/SU(2)$ and expand
the chiral Lagrangian, only keeping terms with up to four pion fields and
space-time derivatives. The resulting Lagrangian takes the form
\begin{equation}
\begin{aligned} \mathscr{L}_{\chi PT} &= \frac{1}{2}\left(\partial_{\mu}\vec{\pi}\right)^{2} 
- \frac{1}{2}M_{\pi}^{2}\vec{\pi}^{2} + C_{1,\chi PT}\left(\vec{\pi}^{2}\right)^{2} 
+ C_{2,\chi PT}\left(\vec{\pi} \cdot \partial_{\mu}\vec{\pi}\right)^{2} 
+ C_{3,\chi PT}\left(\partial_{\mu}\vec{\pi}\right)^{2}\left(\partial_{\nu}\vec{\pi}\right)^{2} \\ 
&\quad + C_{4,\chi PT}\left[\left(\partial_ {\mu}\vec{\pi}\right) \cdot \partial_{\nu}\vec{\pi}\right]^{2} 
+ \mathcal{O}\left(\pi^{6},\partial^{6}\right) \text{ ,} \label{chpt4} \end{aligned}
\end{equation}
where
\begin{equation}
M_{\pi}^{2}=M^{2}+\frac{2\ell_{3}}{f_{\pi}^{2}}M^{4} \label{chpt5}%
\end{equation}
defines the NLO tree-level mass of the pion. The low-energy coupling contants
$C_{i,\chi PT}$, $i=1,\ldots,4$, are given by
\begin{align}
&  C_{1,\chi PT}=-\frac{M^{2}}{8f_{\pi}^{2}}\text{ ,}\label{chpt6}\\
&  C_{2,\chi PT}=\frac{1}{2f_{\pi}^{2}}\text{ ,}\label{chpt7}\\
&  C_{3,\chi PT}=\frac{\ell_{1}}{f_{\pi}^{4}}\text{ ,}\label{chpt8}\\
&  C_{4,\chi PT}=\frac{\ell_{2}}{f_{\pi}^{4}}\text{ ,} \label{chpt9}%
\end{align}
where the LECs are defined in Eq.\ (\ref{app2}).


\section{Low-Energy Limit of the eLSM}

\label{elsm}

\subsection{Mesonic part of the eLSM}

\label{elsmsub1}

The eLSM is a linear sigma model which contains, besides the standard scalar and pseudoscalar 
mesons, also vector and axial-vector mesons \cite{denisnf2,staniold,dick,staninew}. 
All mesonic fields are interpreted as quark-antiquark states, such as e.g.\ found in the relativistic
quark model of Ref.\ \cite{isgur}. This identification is confirmed by the
study of the large-$N_{c}$ behavior \cite{largenc} of masses and widths, as
discussed in Ref.\ \cite{dick}.

Scalar and pseudoscalar mesons are described by the matrix
\begin{equation}
\Phi=\bigl(\sigma_{N}+i\eta_{N}\bigr)T^{0}+\bigl(\vec{a}_{0}+i\vec{\pi
}\bigr)\cdot\vec{T}\text{ ,} \label{elsm1}
\end{equation}
where $T^{0}=\mathbbm{1}_{2\times2}/2$ and $\vec{T}=\vec{\tau}/2$, in which
$\vec{\tau}$ denotes the vector of the Pauli matrices. The quantity $\vec{\pi
}$ describes the pion triplet, while $\eta_{N}$ describes the non-strange
content of the $\eta$ and $\eta^{\prime}$ mesons. Furthermore, the scalar
triplet $\vec{a}_{0}$ is identified with $a_{0}(1450)$ [the alternative
identification with $a_{0}(980)$ turns out to be unfavored
\cite{denisnf2,dick}]. Similarly, the scalar isosinglet $\sigma_{N}$
corresponds to the resonance $f_{0}(1370)$ [also in this case, the assignment
to the light $f_{0}(500)$ is unfavorable, see App.\ \ref{appb2} for further discussion].

Vector and axial-vector mesons are described by the left- and right-handed fields
\begin{align}
L^{\mu}  &  =\bigl(\omega_{N}^{\mu}+f_{1N}^{\mu}\bigr)T^{0}+\bigl(\vec{\rho
}^{\,\mu}+\vec{a}_{1}^{\,\mu}\bigr)\cdot\vec{T}\text{ ,}\label{elsm2}\\
R^{\mu}  &  =\bigl(\omega_{N}^{\mu}-f_{1N}^{\mu}\bigr)T^{0}+\bigl(\vec{\rho
}^{\,\mu}-\vec{a}_{1}^{\,\mu}\bigr)\cdot\vec{T}\text{ ,} \label{elsm3}%
\end{align}
where the vector and axial-vector singlets $\omega_{N}^{\mu}$ and
$f_{1N}^{\mu}$ represent the $\omega(782)$ and $f_{1}(1285)$ mesons,
respectively. In the isotriplet sector, $\vec{\rho}^{\,\mu}$ represents the
vector meson $\rho(770)$ and $\vec{a}_{1}^{\,\mu}$ the axial-vector meson
$a_{1}(1260)$.

The fields (\ref{elsm1}), (\ref{elsm2}), and (\ref{elsm3}) have a well-defined transformation behavior with respect to 
$U(2)_{L}\times U(2)_{R}$ transformations:
\begin{equation}
\Phi\overset{U(2)_{L}\times U(2)_{R}}{\longrightarrow}U_{L}\Phi U_{R}%
^{\dagger}\text{ ,}\quad L^{\mu}\overset{U(2)_{L}\times U(2)_{R}%
}{\longrightarrow}U_{L}L^{\mu}U_{L}^{\dagger}\text{ ,}\quad R^{\mu}%
\overset{U(2)_{L}\times U(2)_{R}}{\longrightarrow}U_{R}R^{\mu}U_{R}^{\dagger
}\text{ .} \label{elsm4}%
\end{equation}
The most general chirally symmetric Lagrangian which contains operators of dimension (up to) four and reproduces
the chiral symmetry breaking pattern found in nature is given by:
\begin{equation}
\begin{aligned} \mathscr{L}_{eLSM} &= \mathrm{Tr}\left\{\left(D^{\mu}\Phi\right)^{\dagger}D_{\mu}\Phi\right\} 
- m_{0}^{2}\mathrm{Tr}\left\{\Phi^{\dagger}\Phi\right\} 
- \lambda_{1} \left(\mathrm{Tr}\left\{\Phi^{\dagger}\Phi\right\}\right)^{2} 
- \lambda_{2}\mathrm{Tr}\left\{\left(\Phi^{\dagger}\Phi\right)^{2}\right\} \\ 
&\quad-\frac{1}{4}\mathrm{Tr}\left\{L^{\mu\nu}L_{\mu\nu} + R^{\mu\nu}R_{\mu\nu}\right\} 
+ \frac{m_{1}^{2}}{2}\mathrm{Tr}\left\{L^{\mu}L_{\mu} + R^{\mu} R_{\mu}\right\} 
+ \mathrm{Tr}\left\{H\left(\Phi^{\dagger} + \Phi\right)\right\} \\
&\quad+ c_{1}\left(\mathrm{det}\Phi - \mathrm{det}\Phi^{\dagger}\right)^2 
+ i\frac{ g_{2}}{2}\left(\mathrm{Tr}\left\{L^{\mu\nu}\left[L_{\mu},L_{\nu}\right]_{-}\right\} 
+ \mathrm{Tr}\left\{R^{\mu\nu}\left[R_{\mu},R_{\nu}\right]_{-}\right\}\right) \\ 
&\quad+ \frac{h_{1}}{2}\mathrm{Tr}\left\{\Phi^{\dagger}\Phi\right\}
\mathrm{Tr}\left\{L^{\mu}L_{\mu} + R^{\mu}R_{\mu}\right\} 
+ h_{2}\mathrm{Tr}\left\{\left|L^{\mu}\Phi\right|^{2} + \left|\Phi R^{\mu}\right|^{2}\right\} 
+ 2h_{3}\mathrm{Tr}\left\{\Phi R^{\mu}\Phi^{\dagger}L_{\mu}\right\} \\ 
&\quad + g_{3}\left(\mathrm{Tr}\left\{L^{\mu}L^{\nu}L_{\mu}L_{\nu}\right\} 
+ \mathrm{Tr}\left\{R^{\mu}R^{\nu}R_{\mu}R_{\nu}\right\}\right) 
+ g_{4}\left(\mathrm{Tr}\left\{L^{\mu}L_{ \mu}L^{\nu}L_{\nu}\right\} 
+ \mathrm{Tr}\left\{R^{\mu}R_{\mu}R^{\nu}R_{\nu}\right\}\right) \\ 
&\quad +g_{5}\mathrm{Tr}\left\{L^{\mu}L_{\mu}\right\}\mathrm{Tr}\left\{R^{\mu}R_{\mu}\right\} 
+ g_{6}\left(\mathrm{Tr}\left\{L^{\mu}L_{\mu}\right\}\mathrm{Tr}\left\{L^{\nu}L_{\nu} \right\} 
+ \mathrm{Tr}\left\{R^{\mu}R_{\mu}\right\}\mathrm{Tr}\left\{R^{\nu}R_{\nu}\right\}\right) \text{ ,} \label{elsm5} 
\end{aligned}
\end{equation}
where $D_\mu \Phi = \partial_\mu \Phi - ig_1 \left( L^\mu \Phi - \Phi R^\mu \right)$, $H=h_{N,0}T^{0}$, 
and $h_{N,0}\sim m_{u}=m_{d}$. 

Explicit breaking of chiral symmetry due to non-vanishing quark masses is affected by the term
\begin{equation}
\mathrm{Tr}\left\{  H\left(  \Phi^{\dagger}+\Phi\right)  \right\}
=h_{N,0}\sigma_{N}\text{ } \label{elsm7}%
\end{equation}
which tilts the potential into the $\sigma_{N}$-direction. The
$U(1)_{A}$ anomaly is incorporated via the term
\begin{equation}
c_{1}\left( \mathrm{det}\Phi-\mathrm{det}\Phi^{\dagger}\right)^{2}\text{ .}
\label{elsm8}%
\end{equation}

Spontaneous breaking of chiral symmetry is induced by a non-vanishing vacuum expectation value
$\phi_{N}\equiv\braket{\sigma_{N}}$ of the $\sigma_{N}$ field. Physical excitations of this field,
corresponding to the $\sigma_{N}$ meson, are described by performing a shift, $\sigma_{N}\longrightarrow\phi_{N}
+\sigma_{N}$, in the Lagrangian (\ref{elsm5}). One then obtains the tree-level masses of the different mesons from
terms quadratic in the fields. In addition, this shift leads to bilinear
terms which mix the axial-vector and pseudoscalar fields, respectively. By shifting the axial-vector fields in an
appropriate way,
\begin{equation}
f_{1N}^{\mu}\longrightarrow f_{1N}^{\mu}+Zw \partial^{\mu}\eta_{N}\text{ , }
\vec{a}_{1}^{\,\mu}\longrightarrow\vec{a}_{1}^{\,\mu}+Zw\partial^{\mu}\vec{\pi}\text{ ,} \label{elsm9}
\end{equation}
with
\begin{equation}
w\equiv \frac{g_{1}\phi_{N}}{m_{a_{1}}^{2}}\text{ ,  }
Z \equiv \bigl(1-g_{1}\phi_{N}w\bigr)^{-\frac{1}{2}}\text{ ,} \label{elsm10}
\end{equation}
the bilinear terms in the Lagrangian can be eliminated [for details, see Refs.\ \cite{dick,staninew}]. 
In addition, we have to redefine the pseudoscalar fields,
\begin{equation}
\eta_{N}\longrightarrow Z\eta_{N}\text{ , }\vec{\pi}\longrightarrow Z\vec{\pi}\;,  \label{elsm11}
\end{equation}
in order to obtain canonically normalized fields. In
the end, the tree-level masses of the mesons read:
\begin{align}
&  m_{\pi}^{2}=\left[  -m_{0}^{2}+\left(  \lambda_{1}+\frac{\lambda_{2}}{2}\right)  \phi_{N}^{2}\right]  
Z^{2}\text{ ,}\nonumber\\
&  m_{\eta_{N}}^{2}=\left[  -m_{0}^{2}+\left(  \lambda_{1}+\frac{\lambda_{2}}{2}\right)  \phi_{N}^{2}
+c_{1}\phi_{N}^{2}\right]  Z^{2}\text{ ,}\nonumber\\
&  m_{a_{0}}^{2}=-m_{0}^{2}+\left(  \lambda_{1}
+\frac{3\lambda_{2}}{2}\right)\phi_{N}^{2}\text{ },\nonumber\\
&  m_{\sigma_{N}}^{2}=-m_{0}^{2}+3\left(  \lambda_{1}
+\frac{\lambda_{2}}{2}\right)  \phi_{N}^{2}\text{ ,}\nonumber\\
&  m_{\omega_{N}}^{2}=m_{\rho}^{2}=m_{1}^{2}
+\frac{1}{2}\left(h_{1}+h_{2}+h_{3}\right)  \phi_{N}^{2}\text{ ,}\nonumber\\
&  m_{f_{1N}}^{2}=m_{a_{1}}^{2}=m_{1}^{2}+\frac{1}{2}\left(h_{1}+h_{2}-h_{3}\right)  \phi_{N}^{2}
+g_{1}^{2}\phi_{N}^{2}\text{ .}
\label{masses}
\end{align}

The parameters $m_{0}^{2},\, m_{1}^{2},\, \lambda_{2},\, g_{1},\, g_{2},\, h_{2},\, h_{3},\,h_{N,0}$, and $c_{1}$ 
of the Lagrangian (\ref{elsm5}) were determined in Ref.\ \cite{dick} through a fit of the tree-level masses (\ref{masses})
as well as several decay widths to experimental data. The large-$N_c$
suppressed parameters $\lambda_1$ and $h_1$ only influence properties of the scalar-isoscalar
mesons and were excluded from this fit. Setting them to zero, the fit allows predictions for the masses of these 
mesons. They turn out to be in the range of masses of the experimentally observed scalar-isoscalar states
$f_0(1370),\, f_0(1500)$, and $f_0(1710)$ \cite{dick}. Note that the fit of Ref.\ \cite{dick} 
was performed for the $N_f=3$ version of the eLSM, while in this work we consider $N_f=2$. Nevertheless,
when $\lambda_1 = h_1 =0$, the terms distinguishing between the cases $N_f=2$ and $N_f=3$ in 
Eqs.\ (\ref{masses}) vanish, such that these relations for the masses 
hold with the identification $c_1^{(N_f=2)} = c_1^{(N_f=3)} \phi_S^2$, where
$\phi_S$ is the strange quark condensate.

The coupling constants $g_{3},$ $g_{4},$ $g_{5},$ $g_{6}$ in the Lagrangian (\ref{elsm5}) describe
four-point interactions between vector mesons. They do not enter masses and
decay widths of mesons at tree-level and thus were not determined in Ref.\ \cite{dick}. As we
shall see in Sec.\ \ref{elsmsub2}, they influence the values of the low-energy coupling constants
because of the mixing of axial-vector and pseudoscalar mesons, Eq.\ (\ref{elsm9}), 
which gives rise to a four-pion term. While the constants $g_{5}$ and $g_{6}$ can be dismissed
in virtue of large-$N_{c}$ considerations, the constants $g_{3}$ are $g_{4}$
are not expected to be small.

\subsection{Determination of the low-energy effective action of the eLSM}

\label{elsmsub2}

In this subsection, we compute the low-energy effective action of the eLSM by successively integrating
out all fields except for the pions in the vacuum-to-vacuum transition amplitude
$\braket{f, \infty | f, -\infty}$, where $f=\left\{  \sigma_{N},\eta_{N},\vec{a}_{0},\vec{\pi},\omega_{N}^{\mu},
f_{1N}^{\mu},\vec{\rho}^{\,\mu},\vec{a}_{1}^{\,\mu}\right\} $. This transition amplitude can be written as a 
functional integral over all fields
\begin{equation}
\begin{aligned} \braket{f,\infty | f, -\infty} &= \mathcal{N}\int\mathscr{D}f\; 
\exp\biggl(i\int\mathrm{d}^{4}x \; \mathscr{L}_{eLSM} \biggr) \text{ ,} \label{elsm18} \end{aligned}
\end{equation}
where $\mathcal{N}$ is a normalization constant. Our aim is to obtain a
Lagrangian which contains only pions and can then be compared to ChPT. Hence, we
integrate out the heavy mesonic fields $H\equiv\{\sigma_{N},\,\eta_{N},\,\vec{a}_{0},\,\omega_{N}^{\mu},\,
f_{1N}^{\mu},\,\vec{\rho}^{\,\mu},\, \vec{a}_{1}^{\,\mu}\}$. In general, this is a formidable task, since
the various interaction structures couple the functional integrations over different fields in Eq.\ (\ref{elsm18}). 
Furthermore, there are cubic and quartic (self-)interaction terms of the heavy fields, which prevent a
straightforward analytic solution of the respective functional integral. 

Nevertheless, if we restrict the comparison of ChPT with the low-energy effective action of the eLSM
to {\it tree-level\/} and to {\it fourth order in powers of derivatives of the pion fields\/}, it is possible to make progress 
by purely 
analytical means. We first observe that the Lagrangian (\ref{elsm5}) contains the following type of interaction terms:
\begin{itemize}
\item [(1)] terms containing three or four heavy fields, but no pion field,
$\Gamma^{(3)}_H \equiv H_i H_j H_k$ and $\Gamma^{(4)}_H \equiv
H_{i}H_{j}H_{k} H_l$, where $H_i,\, H_j,\, H_k,\, H_l$, are heavy fields,
\item [(2)] terms containing one pion and two or three heavy fields,
$\Gamma^{(3)}_\pi \equiv H_i H_j \pi$ and $\Gamma^{(4)}_\pi \equiv
H_{i}H_{j}H_{k}\pi$, 
\item [(3)] terms containing two pion fields and one heavy field, 
$\Gamma^{(3)}_{\pi\pi} \equiv H_{i}\pi\pi$,
\item [(4)] terms containing two pion fields and two heavy fields, 
$\Gamma^{(4)}_{\pi\pi} \equiv H_{i}H_j\pi\pi$,
\item [(5)] terms containing three pion fields and one heavy field $\Gamma^{(4)}_{\pi\pi\pi} \equiv H_{i}\pi\pi\pi$,
and
\item[(6)] terms containing four pion fields $\Gamma^{(4)}_{\pi\pi\pi\pi} \equiv \pi\pi\pi\pi$.
\end{itemize}
Under the above assumptions, we may now neglect all terms except those of type (3) and (6).
This can be proved as follows. 
Because of our assumption to consider only terms up to fourth order in powers of derivatives of the pion fields, 
we need to combine the different types of vertices in a way which generates four-pion interaction terms when
integrating out the heavy fields.
However, one can convince oneself via a simple graphical analysis that most of these combinations then contain
loops of heavy fields. By our assumption to consider only tree-level contributions to the low-energy
effective action, these can therefore be neglected.
The only way to generate a tree-level contribution is to combine two vertices of type (3) 
with the same heavy field $H_i$. When integrating out the latter, this generates a diagram where the two vertices
of type (3) are connected by a propagator for the field $H_i$.
Other than that, the only other terms that contribute at tree-level are those of type (6).

After these considerations, the only terms of relevance in the Lagrangian (\ref{elsm5}) are
\begin{equation}
\mathscr{L}_{eLSM} = \mathscr{L}_{\pi}+\mathscr{L}_{\sigma_{N}\pi}+\mathscr{L}_{\rho \pi}
+ \ldots \text{,}
\label{elsmseries}%
\end{equation}
where
\begin{align}
\mathscr{L}_{\pi}  &  =\frac{1}{2}\left(  \partial_{\mu}\vec{\pi}\right)^{2}
-\frac{1}{2}m_{\pi}^{2}\vec{\pi}^{2}+\frac{g_{1}^{2}}{2}w^{2}Z^{4}
\left(  \vec{\pi}\cdot\partial_{\mu}\vec{\pi}\right)^{2}
-\frac{1}{4}\left(  \lambda_{1}+\frac{\lambda_{2}}{2}\right)  Z^{4}
\left(\vec{\pi}^{2}\right)  ^{2}+\frac{1}{4}\left(  h_{1}+h_{2}-h_{3}\right)
w^{2}Z^{4}\vec{\pi}^{2}\left(  \partial_{\mu}\vec{\pi}\right)  ^{2}\nonumber\\
&  \quad +\frac{h_{3}}{2}w^{2}Z^{4}\left(  \vec{\pi}\times\partial_{\mu}\vec{\pi}\right)^{2}
+\left(  -\frac{g_{3}}{4}+\frac{g_{4}}{4}+\frac{g_{5}}{4}
+\frac{g_{6}}{2}\right)  w^{4}Z^{4}\left(  \partial_{\mu}\vec{\pi}\right)^{2}
\left(  \partial_{\nu}\vec{\pi}\right)^{2}+\frac{g_{3}}{2}w^{4}Z^{4}\left(  \partial^{\mu}\vec{\pi}\right)
\cdot\left(  \partial^{\nu}\vec{\pi}\right)  \left(  \partial_{\mu}\vec{\pi}\right)  
\cdot\partial_{\nu}\vec{\pi}
\label{pionpart}
\end{align}
contains the kinetic and mass contributions of
the $\pi$-field as well as all types of four-pion interaction terms, and
\begin{align}
\mathscr{L}_{\sigma_{N}\pi}  &  =\frac{1}{2}\left(  \partial_{\mu}\sigma_{N}\right)^{2}
-\frac{1}{2}m_{\sigma_{N}}^{2}\sigma_{N}^{2}-\left(\lambda_{1}+\frac{\lambda_{2}}{2}\right)  
\phi_{N}Z^{2}\sigma_{N}\vec{\pi}^{2}+g_{1}wZ^{2}\left(  \partial^{\mu}\sigma_{N}\right)  
\left(  \partial_{\mu}\vec{\pi}\right)  \cdot\vec{\pi}\nonumber\\
&  \quad+\left.  \left. \biggl\{\biggl[g_{1}^{2}\phi_{N} + \left(  h_{1}+h_{2}-h_{3}\right)  \frac{\phi_{N}}{2}\right]  w^{2}Z^{2}
-g_{1}wZ^{2}\right\}  \sigma_{N}\left(  \partial_{\mu}\vec{\pi}\right)^{2}\text{ ,} \label{sigmanpart}%
\end{align}
as well as 
\begin{align}
\mathscr{L}_{\rho\pi}  &  =-\frac{1}{4}\vec{\rho}^{\,\mu\nu} \cdot\vec{\rho}_{\mu\nu}
+\frac{1}{2}m_{\rho}^{2}\vec{\rho}_{\mu}^{\,2}-\frac{\xi_{\rho}}{2}
\left(  \partial_{\mu}\vec{\rho}^{\,\mu}\right)^{2}+g_{2}w^{2}Z^{2}\left(  \partial^{\mu}\vec{\rho}^{\,\nu}\right)
\cdot\left(  \partial_{\nu}\vec{\pi}\times\partial_{\mu}\vec{\pi}\right)
\nonumber\\
&  \quad+\left[  \left(  g_{1}^{2}\phi_{N}-h_{3}\phi_{N}\right)  wZ^{2}-  g_{1}Z^{2}\right]  \vec{\rho}^{\,\mu}\cdot
\left(  \partial_{\mu}\vec{\pi}\times\vec{\pi}\right)   \label{rhopart}%
\end{align}
contain the kinetic and mass terms as well as the terms linear in the $\sigma_N$ and the $\rho$ field, respectively.
Note that we added a St\"uckelberg term 
$\xi_\rho (\partial_\mu \vec{\rho}^{\,\mu})^2/2$ to
the Lagrangian in order to make the inverse $\rho$-meson propagator invertible, cf.\ App.\ \ref{appa2}.

The remaining terms [denoted by the ellipsis in Eq.\ (\ref{elsmseries})] contain the kinetic and mass terms 
for the other heavy fields $H=\eta_{N},\,\vec{a}_{0},\,\omega_{N}^{\mu},\,f_{1N}^{\mu},\,\vec{a}_{1}^{\,\mu}$ 
as well as their interaction terms with pions. However, they do not contain any term of type (3), cf.\ App.\
{\ref{appa2}}, and therefore can be neglected within our approximation scheme. 

Since the Lagrangian (\ref{elsmseries}) contains at most quadratic terms in the heavy fields, the
integration over the latter in the functional integral (\ref{elsm18}) is of (shifted) Gaussian type and
can therefore be performed analytically. 
Using Eq.\ (\ref{elsmseries}) the functional integral (\ref{elsm18}) takes the form:
\begin{equation}
\braket{f,\infty | f, -\infty}=\mathcal{N}\int\mathscr{D}\pi\;\exp\left(  i\int\mathrm{d}^{4}x\;\mathscr{L}_{\pi}\right)
\prod\limits_{H=\sigma_N,\rho}I_{H}[\pi]\text{ ,} \label{elsm21}%
\end{equation}
where
\begin{equation}
I_{\sigma_{N}}[\pi]=\int\mathscr{D}\sigma_{N}\;\exp\left(
i\int\mathrm{d}^{4}x\;\mathscr{L}_{\sigma_{N}\pi}\right)  \text{
,}\quad\quad I_{\rho}[\pi]=\int\mathscr{D}\sigma_{N}\;
\exp\left(  i\int\mathrm{d}^{4}x\;\mathscr{L}_{\rho\pi}\right)  \text{ ,} \label{elsm23}%
\end{equation}
with $\mathscr{L}_{\sigma_{N}\pi}$ and $\mathscr{L}_{\rho\pi}$ are given by Eqs.\ (\ref{sigmanpart}) and 
(\ref{rhopart}), respectively. Since both functional integrals decouple, the order in which we integrate out
these fields is irrelevant. We start with $\sigma_{N}$:
\begin{align}
I_{\sigma_{N}}[\pi]  &  =\int\mathscr{D}\sigma_{N}\;\exp\left[
-\frac{i}{2}\int\mathrm{d}^{4}x\mathrm{d}^{4}y\;\sigma_{N}(x)
\mathscr{O}_{\sigma_{N}}(x,y)\sigma_{N}(y)
+i\int\mathrm{d}^{4}xJ_{\sigma_{N}\pi}(x)\sigma_{N}(x)\right] \nonumber\\
&  =\mathcal{N}_{\sigma_{N}}\exp\left[  \frac{i}{2}\int\mathrm{d}^{4}x\mathrm{d}^{4}y\;J_{\sigma_{N}\pi}(x)
\mathscr{O}_{\sigma_{N}}^{-1}(x,y)J_{\sigma_{N}\pi}(y)\right] \text{ ,}
\label{elsm24}
\end{align}
where
\begin{align}
\mathscr{O}_{\sigma_{N}}^{-1}(x,y)  &  =\left(  \Box_{x}+m_{\sigma_{N}}^{2}\right)^{-1}\delta^{(4)}(x-y) 
=\frac{1}{m_{\sigma_{N}}^{2}}\sum\limits_{n=0}^{\infty}(-1)^{n}
\left(\frac{\Box_{x}}{m_{\sigma_{N}}^{2}}\right)^{n}\delta^{(4)}(x-y)\label{elsm25}
\end{align}
and
\begin{equation}
J_{\sigma_{N}\pi}(x)=c_{1,\sigma_{N}}\vec{\pi}^{2}
+c_{2,\sigma_{N}}\left(  \partial_{\mu}\vec{\pi}\right)^{2}\text{ ,} \label{elsm26}
\end{equation}
with the coefficients 
\begin{align}
c_{1,\sigma_{N}}  &  =g_{1}wZ^{2}m_{\pi}^{2}-\left(  \lambda_{1}+\frac{\lambda_{2}}{2}\right)  
\phi_{N}Z^{2}\text{ ,}\label{elsm27}\\
c_{2,\sigma_{N}}  &  =\left[  g_{1}^{2}\phi_{N}+\left(  h_{1}+h_{2}-h_{3}\right)  \frac{\phi_{N}}{2}\right]  
w^{2}Z^{2}-2g_{1}wZ^{2}\text{ .}
\label{elsm28}
\end{align}
Expanding the sum in Eq.\ (\ref{elsm25}) to order $n=2$, neglecting terms of higher than fourth order in derivatives
of the pion fields, and using the equation of motion
of the free pion field, Eq.\ (\ref{elsm24}) can be written as
\begin{align}
I_{\sigma_{N}}[\pi]  &  =\mathcal{N}_{\sigma_{N}}\exp\left(
i\int\mathrm{d}^{4}x\;\left\{  \left[  \frac{c_{1,\sigma_{N}}^{2}}{2m_{\sigma_{N}}^{2}}
\left(  1-\frac{4m_{\pi}^{4}}{m_{\sigma_{N}}^{4}}\right)  
+\frac{c_{1,\sigma_{N}}c_{2,\sigma_{N}}m_{\pi}^{2}}{m_{\sigma_{N}}^{2}}
\left(  1+\frac{2m_{\pi}^{2}}{m_{\sigma_{N}}^{2}}\right)  \right]  \left(  \vec{\pi}^{2}\right)^{2}\right.  \right. \nonumber \\
& \hspace*{3.3cm} +\left[  \frac{2c_{1,\sigma_{N}}^{2}}{m_{\sigma_{N}}^{4}}
\left(  1+\frac{4m_{\pi}^{2}}{m_{\sigma_{N}}^{2}}\right)  
-\frac{2c_{1,\sigma_{N}}c_{2,\sigma_{N}}}{m_{\sigma_{N}}^{2}}\left(  1
+\frac{2m_{\pi}^{2}}{m_{\sigma_{N}}^{2}}\right)  \right]  
\left(  \vec{\pi}\cdot\partial_{\mu}\vec{\pi}\right)^{2}\nonumber\\
&  \hspace*{3.3cm}+\Biggl.\Biggl.\left[  \frac{c_{2,\sigma_{N}}^{2}}{2m_{\sigma_{N}}^{2}}
-\frac{2c_{1,\sigma_{N}}c_{2,\sigma_{N}}}{m_{\sigma_{N}}^{4}}
+\frac{2c_{1,\sigma_{N}}^{2}}{m_{\sigma_{N}}^{6}}\right]  \left(
\partial_{\mu}\vec{\pi}\right)  ^{2}\left(  \partial_{\nu}\vec{\pi}\right)^{2}\Biggr\}\Biggr)\text{ .} \label{elsm29}
\end{align}

We now turn to the contribution of the $\rho$-meson:
\begin{align}
I_{\rho}[\pi]  &  =\int\mathscr{D}\rho\;\exp
\biggl[\frac{1}{2}\int\mathrm{d}^{4}x\mathrm{d}^{4}y\;\vec{\rho}_{\mu}(x)\cdot
\mathscr{O}_{\rho}^{\mu\nu}(x,y)\vec{\rho}_{\nu
}(y)+i\int\mathrm{d}^{4}x\;\vec{J}_{\rho\pi}^{\mu}(x)\cdot
\vec{\rho}_{\mu}(x)\biggr]\nonumber\\
&  =\mathcal{N}_{\rho}\exp\biggl[-\frac{i}{2}\int\mathrm{d}^{4}x\mathrm{d}^{4}y\;
\vec{J}_{\rho \pi,\mu}(x)\cdot\mathscr{O}_{\rho}^{\mu\nu,-1}(x,y)\vec{J}_{\rho \pi,\nu}(y)
\biggr]\text{ ,} \label{elsm30}
\end{align}
where
\[
\mathscr{O}_{\rho}^{\mu\nu,-1}(x,y)=\left[  \frac{g^{\mu\nu}}{\Box_{x}+m_{\rho}^{2}}
-\frac{1-\xi_{\rho}}{\left(  \xi_{\rho} \Box_{x}+m_{\rho}^{2}\right)  \left(  \Box_{x}+m_{\rho}^{2}\right)  }
\partial_{x}^{\mu}\partial_{x}^{\nu}\right]  \delta^{(4)}(x-y)
\]
is the propagator of the $\rho$-meson and
\begin{equation}
\vec{J}_{\rho\pi}^{\mu}(x)=c_{1,\rho}\left[  \left(
\partial^{\mu}\vec{\pi}\right)  \times\vec{\pi}\right]  -c_{2,\rho}
\left[  \left(  \partial^{\mu}\partial^{\nu}\vec{\pi}\right)
\times\partial_{\nu}\vec{\pi}\right]  \text{ ,} \label{elsm32}
\end{equation}
with the coefficients 
\begin{align}
&  c_{1,\rho}=\left(  g_{1}^{2}-h_{3}\right)  \phi_{N}wZ^{2}-g_{1}
Z^{2}+g_{2}w^{2}Z^{2}m_{\pi}^{2}\text{ ,}\label{elsm33}\\
&  c_{2,\rho}=g_{2}w^{2}Z^{2}\text{ .} \label{elsm34}
\end{align}
At this point, for the sake of simplicity we choose $\xi_{\rho}=1$, which eliminates the term
proportional to $\partial_{x}^{\mu}\partial_{x}^{\nu}$ in the inverse propagator. This does not influence
our results, as one can show that this term results in
four-pion interaction terms with six or more space-time derivatives, which we neglect in our treatment. 
Then, the inverse operator in Eq.\ (\ref{elsm30}) simplifies and the functional integral with respect to
$\rho^{\mu}$ can finally be written as
\begin{align}
I_{\rho}[\pi]  &  =\mathcal{N}_{\rho} \exp\left(-i\int\mathrm{d}^{4}x\;\left\{\left(
\frac{c_{1,\rho}^{2}m_{\pi}^{2}}{2m_{\rho}^{2}}
-\frac{c_{1,\rho}c_{2,\rho}m_{\pi}^{4}}{m_{\rho}^{2}}\right)  \left(  \vec{\pi}^{2}\right)  ^{2} 
-\left(  \frac{3c_{1,\rho}^{2}}{2m_{\rho}^{2}}
-\frac{3c_{1,\rho}c_{2,\rho}m_{\pi}^{2}}{m_{\rho}^{2}}\right)  \left(  \vec{\pi}\cdot\partial_{\mu}\vec{\pi}\right)^{2}
\right.\right.\nonumber\\
&  \quad\quad\quad\quad\;\;+\left.\left.\left(\frac{c_{1,\rho}^{2}}{m_{\rho}^{4}}
+\frac{c_{1,\rho}c_{2,\rho}}{m_{\rho}^{2}}\right)  
\left(  \partial_{\mu}\vec{\pi}\right)^{2}\left(  \partial_{\nu}\vec{\pi}\right)^{2}
-\left(  \frac{c_{1,\rho}^{2}}{m_{\rho}^{4}}+\frac{c_{1,\rho}c_{2,\rho}}{m_{\rho}^{2}}\right) 
 \left[ \left(  \partial_{\mu}\vec{\pi}\right)  \cdot\partial_{\nu}\vec{\pi}\right]^2
\right\}\right)\text{ .} \label{elsm35}
\end{align}

In order to obtain the tree-level effective action of the eLSM, we have to
insert Eqs.\ (\ref{elsm29}) and (\ref{elsm35}) into Eq.\ (\ref{elsm21}). The
resulting tree-level effective Lagrangian has then exactly the same form as
Eq.\ (\ref{chpt4}) obtained from ChPT:
\begin{align}
\mathscr{L}_{eLSM,eff}[\vec{\pi}]  &  =\frac{1}{2}\left(  \partial_{\mu}%
\vec{\pi}\right)  ^{2}-\frac{1}{2}m_{\pi}^{2}\vec{\pi}^{2}%
+C_{1,eLSM}\left(  \vec{\pi}^{2}\right)  ^{2}+C_{2,eLSM}\left[\left(  \partial_{\mu
}\vec{\pi}\right)\cdot\vec{\pi}\right]^{2}+C_{3,eLSM}\left(  \partial_{\mu}\vec
{\pi}\right)  ^{2}\left(  \partial_{\nu}\vec{\pi}\right)  ^{2}\nonumber\\
&  \quad+C_{4,eLSM}\left[  \left(  \partial_{\mu}\vec{\pi}\right)
\cdot\partial_{\nu}\vec{\pi}\right]  ^{2}\text{ ,} \label{elsm36}%
\end{align}
where the low-energy coupling constants $C_{i,eLSM}$, $i=1,\ldots,4$, are  
functions of the parameters of the eLSM Lagrangian (\ref{elsm5}):
\begin{align}
&  C_{1,eLSM}=\frac{Z^{4}}{4}\left[  \left(  h_{1}+h_{2}+h_{3}\right)
w^{2}m_{\pi}^{2}-\left(  \lambda_{1}+\frac{\lambda_{2}}{2}\right)
\right]  +\frac{c_{1,\sigma_{N}}^{2}}{2m_{\sigma_{N}}^{2}}\left(
1-\frac{4m_{\pi}^{4}}{m_{\sigma_{N}}^{4}}\right)  
\nonumber\\
&  \quad\quad\quad\quad\quad   +\frac{c_{1,\sigma_{N}}c_{2,\sigma_{N}}m_{\pi}^{2}}{m_{\sigma_{N}}^{2}}
  \left(  1+\frac{2m_{\pi}^{2}}{m_{\sigma_{N}}^{2}}\right)  
-\frac{c_{1,\rho}^{2}m_{\pi}^{2}}{2m_{\rho}^{2}}+\frac{c_{1,\rho}c_{2,\rho}m_{\pi}^{4}}{m_{\rho}^{2}}\text{ ,}
\label{elsmchpt1}\\
&  C_{2,eLSM}=\frac{1}{2}\left(  g_{1}^{2}-h_{1}-h_{2}-2h_{3}\right)
w^{2}Z^{4}+\frac{2c_{1,\sigma_{N}}^{2}}{m_{\sigma_{N}}^{4}}\left(
1+\frac{4m_{\pi}^{2}}{m_{\sigma_{N}}^{2}}\right)  -\frac{2c_{1,\sigma_{N}}c_{2,\sigma_{N}}}{m_{\sigma_{N}}^{2}}
\left(  1+\frac{2m_{\pi}^{2}}{m_{\sigma_{N}}^{2}}\right) \nonumber\\
&  \quad\quad\quad\quad\quad+\frac{3c_{1,\rho}^{2}}{2m_{\rho}^{2}}
-\frac{3c_{1,\rho}c_{2,\rho}m_{\pi}^{2}}{m_{\rho}^{2}}\text{ ,}\label{elsmchpt2}\\
&  C_{3,eLSM}=\frac{1}{4}\left(-g_{3}+g_{4}+g_{5}
+2g_{6}\right)  w^{4}Z^{4}+\frac{c_{2,\sigma_{N}}^{2}}{2m_{\sigma_{N}}^{2}}
-\frac{2c_{1,\sigma_{N}}c_{2,\sigma_{N}}}{m_{\sigma_{N}}^{4}}
+\frac{2c_{1,\sigma_{N}}^{2}}{m_{\sigma_{N}}^{6}}-\frac{c_{1,\rho}^{2}}{m_{\rho}^{4}}
-\frac{c_{1,\rho}c_{2,\rho}}{m_{\rho}^{2}}\text{ ,}\label{elsmchpt3}\\
&  C_{4,eLSM}=\frac{g_{3}}{2}w^{4}Z^{4}+\frac{c_{1,\rho}^{2}}{m_{\rho}^{4}}
+\frac{c_{1,\rho}c_{2,\rho}}{m_{\rho}^{2}}\text{ .} \label{elsmchpt4}%
\end{align}
These expressions represent the main result of this paper. 
Note that they already contain contributions from the $\sigma_N$- and the $\rho$-meson at tree-level.
In other approaches without these meson degrees of freedom, e.g.\ ChPT, such contributions only enter at 
higher-loop order.

\section{Results}

\label{res}

In this section we determine the numerical values for the low-energy coupling constants
$C_{i,\chi PT}$ and $C_{i,eLSM}$, $i=1,\ldots,4$, and compare them with each other.

\subsection{ChPT}

\label{ressub1}

In ChPT at NLO the LECs are functions of the energy scale $\mu$ and are
denoted as $\ell_{i}(\mu).$ They are related to the usually quoted $\mu
$-independent quantities $\bar{\ell}_{i}$ through the equation:
\begin{equation}
\bar{\ell}_{i}=\frac{32\pi^{2}}{\gamma_{i}}\ell_{i}(\mu)-\ln\frac{M_{\pi}^{2}}{\mu^2}\text{ ,}%
\end{equation}
where $\gamma_{1}=1/3$, $\gamma_{2}=2/3$, and
$\gamma_{3}=-1/2$, see Ref.\ \cite{gale}. The numerical values of the $\bar{\ell}_{i}$
are given in Ref.\ \cite{bijnens}. For the pion mass we use $M_\pi= (139.57018 \pm 0.00035)$ MeV and for 
the renormalization scale $\mu=770$ MeV, resulting in the following values for the 
$\ell_i(\mu=770 \textrm{ MeV})$: 
\begin{align}
&  \ell_{1}=\left(  -4.03\pm0.63\right)  \cdot10^{-3}\text{ ,}\label{l1exp}\\
&  \ell_{2}=\left(  1.87\pm0.21\right)  \cdot10^{-3}\text{ ,}\label{l2exp}\\
&  \ell_{3}=\left(  0.8\pm3.9\right)  \cdot10^{-3}\text{ .} \label{l3exp}
\end{align}
In this way the mass parameter $M$, which is defined by the NLO tree-level
mass of the pion, Eq.\ (\ref{chpt5}), reads:
\begin{equation}
M=(139.3\pm1.2)\text{ MeV}\text{ ,}
\end{equation}
Upon using $f_{\pi}=\left(  92.2\pm0.1\right)  $ MeV \cite{pdg} one obtains: 
\begin{align}
&  C_{1,\chi PT}=-0.29\pm0.34\text{ ,}\label{c1chpt}\\
&  C_{2,\chi PT}=(5.882\pm0.013)\cdot10^{-5}\text{ MeV}^{-2}\text{
,}\label{c2chpt}\\
&  C_{3,\chi PT}=(-5.57\pm0.88)\cdot10^{-11}\text{ MeV}^{-4}\text{
,}\label{c3chpt}\\
&  C_{4,\chi PT}=(2.58\pm0.29)\cdot10^{-11}\text{ MeV}^{-4}\text{ .}
\label{c4chpt}
\end{align}

\subsection{eLSM}

\label{ressub2}

The parameters of the eLSM were determined in Ref.\ \cite{dick} through a fit
to experimental data. We only quote those of relevance for the following:
\begin{align}
g_{1}  &  =5.843\pm0.018\text{ ,}\\
g_{2}  &  =3.02\pm0.23\text{ ,}\\
h_{2}  &  =9.88\pm0.66\text{ ,}\\
h_{3}  &  =4.87\pm0.086\text{ }\\
\lambda_{2}  &  =68.297\pm0.044\text{ ,}\\
m_{0}^{2}  &  =\left(-0.91825\pm0.00064\right)\text{ GeV}^{2}\\
m_{1}^{2}  &  =\left(0.4135\pm0.015\right)\text{ GeV}^{2}\;.
\end{align}
We also set $h_{1}=\lambda_{1}=g_{5}=g_{6}=0$, which are
suppressed in the large-$N_c$ limit \cite{largenc}. The minimum of the
potential is at $\phi_N  =\left( 164.6 \pm 0.1 \right)$ MeV. As a consequence, we obtain
\begin{align}
& m_\pi = \left(  141.0\pm5.8\right)  \text{ MeV}\text{ ,}\\
& m_{\sigma_{N}}=1362.7 \text{ MeV}\text{ ,}\\
& m_{\rho}=\left(  783.1\pm7.0\right) \text{ MeV}\text{ ,}\\
& m_{a_{1}}=\left(1185.6\pm5.6\right)\text{ MeV}\text{ ,}\\
& w=(6.838\pm0.072)\cdot10^{-4}\text{ MeV}^{-1}\text{ ,}\\
& Z=1.71\pm0.18\text{ .}
\end{align}
The four coefficients $c_{i,\sigma_{N}}$ and $c_{i,\rho}$, $i=1,2$, defined
in Eqs.\ (\ref{elsm27}), (\ref{elsm28}), (\ref{elsm33}), and (\ref{elsm34}), are then:
\begin{align}
&  c_{1,\sigma_{N}}=(-1.62\pm0.34)\cdot10^{4}\text{ MeV}\text{ ,}\\
&  c_{2,\sigma_{N}}=(-0.0151\pm0.0032)\text{ MeV}^{-1}\text{ ,}\\
&  c_{1,\rho}=-7.4\pm1.6\text{ ,}\\
&  c_{2,\rho}=(4.13\pm0.93)\cdot10^{-6}\text{ MeV}^{-2}\text{ .}
\end{align}
Using these values, the low-energy coupling constants of the eLSM, Eqs.\
(\ref{elsmchpt1}) -- (\ref{elsmchpt4}), are:
\begin{align}
&  C_{1,eLSM}=-0.268\pm0.021\text{ ,}\label{c1elsm}\\
&  C_{2,eLSM}=(5.399\pm0.081)\cdot10^{-5}\text{ MeV}^{-2}\text{ ,}%
\label{c2elsm}\\
&  C_{3,eLSM}=(-9.30\pm0.59)\cdot10^{-11}\text{MeV}^{-4}+\left(
-\frac{g_{3}}{4}+\frac{g_{4}}{4}\right)  w^{4}Z^{4}\text{ ,}\label{c3elsm}\\
&  C_{4,eLSM}=(9.45\pm0.59)\cdot10^{-11}\text{MeV}^{-4}+\text{ }\frac{g_{3}}{2}w^{4}Z^{4}\text{ ,}\label{c4elsm}%
\end{align}
where the errors are calculated using the standard procedure associated
to the $\chi^{2}$ minimization described in Ref.\ \cite{dick}. The following
comments are in order:
\begin{itemize}
\item [1)] The constant $C_{1,eLSM}$ turns out to be in good agreement with the ChPT
result. The eLSM value has, quite remarkably, an even smaller error than $C_{1,\chi PT}$. 
However, this does not mean that we can determine $\ell_3$ to better precision than
ChPT. Naively one would think that Eq.\ (\ref{chpt5}) allows us to express $\ell_3$ as a function
of $M^2$, which, by Eq.\ (\ref{chpt6}), is linearly related to $C_{1,\chi PT}$. We could now
replace $C_{1,\chi PT}$ with $C_{1,eLSM}$ and hope to obtain a smaller error for $\ell_3$ than
ChPT. This, however, does not work: we obtain $\ell_3 =  (23 \pm 28) \cdot 10^{-3}$, 
i.e., although the value is consistent with the one quoted in Eq.\ (\ref{l3exp}) it has an error which is about one
order of magnitude larger.
The reason is that the mass difference $M_{\pi}^{2}-M^{2}$ has a larger error which influences this
way of determining $\ell_3$.
\item [2)] The quantity $C_{2,eLSM}$ is a few standard deviations off the NLO ChPT
value. However, if we consider $C_{2,eLSM} (2 f_\pi^2)$, with the value for $f_\pi$ as
given by the fit of Ref.\ \cite{dick}, $f_\pi = (96.3 \pm 0.7)$ MeV, we obtain
$C_{2,eLSM} (2f_{\pi}^{2})=1.00129 \pm 0.00012$, i.e., almost exactly equal to unity (although the error
is about a factor of 10 smaller than the deviation from unity).
It is interesting to list the five terms contributing to the
right-hand side of Eq.\ (\ref{elsmchpt2}) separately:
\begin{equation}
C_{2,eLSM} \, (2f_{\pi}^{2})=0.53775 + 2.93915 - 4.98863 + 2.45817 + 0.05484 =1.00129\;.
\end{equation}
From this we conclude that the result $C_{2,eLSM} (2f_{\pi}^{2})\simeq1$ is actually due to
nontrivial cancellations.
\item [3)] The quantities $C_{3,eLSM}$ and $C_{4,eLSM}$ cannot be uniquely determined because
the constants $g_{3}$ and $g_{4}$ were not determined in the fit of Ref.\
\cite{dick}. However, we can estimate their values, by replacing the left-hand sides
of Eqs.\ (\ref{c3elsm}) and (\ref{c4elsm}) by the ChPT values (\ref{c3chpt}) and (\ref{c4chpt}) 
and then solving for $g_3$ and $g_4$. We obtain
\begin{align}
g_{3}  &  =-74\pm32\text{ ,}\\
g_{4}  &  = 6\pm52\text{ .}%
\end{align}
Although the errors are large, the values are of a natural order of magnitude. 
In turn, it means that the terms proportional to $g_{3}$ and
$g_{4}$ are expected to affect $\pi\pi$ scattering.
\end{itemize}

\section{Conclusions and outlook}

\label{con}

In this work, we have presented a low-energy study of the eLSM with two quark
flavors. We have integrated out all heavy mesons in the functional integral representation of the
vacuum-to-vacuum transition amplitude and kept only terms contributing at tree-level and
up to fourth order in powers of derivatives of the pion field. In this way, we have obtained a low-energy effective action
which contains only pions. We have mapped this effective action to 
that of ChPT by choosing a definite coset representative and expanding the latter to fourth
order in powers of derivatives of the pion fields.
This allowed us to compare the coefficients of the various terms, here termed low-energy coupling constants, 
in the low-energy effective action of the eLSM with the corresponding ones of ChPT. 

The low-energy coupling constant $C_{1, \chi PT}$ is related to the LEC $\ell_3$, cf.\  Eqs.\ (\ref{chpt5}) and
(\ref{chpt6}), while $C_{2, \chi PT} = 1/(2 f_\pi^2)$. On the other hand
$C_{1,eLSM}$ and $C_{2,eLSM}$ were determined from the fit of Ref.\
\cite{dick}. We found reasonable agreement between $C_{1,eLSM}$ and $C_{1, \chi PT}$, while the numerical values
of $C_{2,eLSM}$ and $C_{2, \chi PT}$ differ by a few standard deviations.
However, if we consider the dimensionless quantity $C_{2,eLSM}(2 f_\pi^2)$, with
$f_\pi$ from the fit of Ref.\ \cite{dick}, we obtain within about 0.13\% the
value $C_{2,eLSM}(2 f_\pi^2)=1$. 

A direct comparison of the low-energy coupling constants $C_{3,eLSM}$ and $C_{4,eLSM}$ with the corresponding
ones in ChPT was at present not possible because the coupling constants $g_3$ and
$g_4$ have not yet been determined. In view of this we reverted the argument and obtained an estimate for
these couplings constants by equating $C_{3,eLSM} \equiv C_{3,\chi PT}$ and $C_{4,eLSM} \equiv C_{4,\chi PT}$.
We obtained values which are of a natural order of magnitude. It would be interesting to study $\pi\pi$ scattering
within the eLSM to confirm the values obtained here.

In conclusion, we confirmed the validity of the eLSM as an effective hadronic model by showing 
that its low-energy limit correctly reproduces the low-energy coupling constants of ChPT to NLO. A necessary
ingredient proved to be the inclusion of (axial-)vector degrees of freedom. 
In App.\ \ref{appb1} we corroborate this conclusion by studying a scenario
without (axial-)vector mesons. Let us also repeat the main conclusion of Ref.\ \cite{dick},
namely that the scalar quark-antiquark states lie above 1 GeV in mass and in particular that the chiral partner of the 
pion has to be identified with the $f_0(1370)$ resonance. In App.\ \ref{appb2} we study an alternative scenario
where the scalar quark-antiquark states are identified with resonances below 1 GeV in mass. In this case, we show that
the low-energy coupling constants of the eLSM disagree with those of ChPT.

At present, our conclusions hold at tree-level. Therefore, a mandatory future project is to compute 
loop corrections to the low-energy coupling constants for the eLSM in order to confirm that the eLSM has
the same low-energy effective action as QCD.

\bigskip

\textbf{Acknowledgments}: The authors thank D.D.\ Dietrich, J.\ Eser, B.\ Kubis, D.\ Parganlija, J.M.\ Pawlowski, 
H.\ van Hees, and Gy.\ Wolf for useful discussions.

\appendix

\section{Lagrangians}

\label{appa}

\subsection{ChPT}

\label{appa1}

At LO, the ChPT Lagrangian is given by
\begin{equation}
\mathscr{L}_{2} = \frac{f^{2}_{\pi}}{4}\mathrm{Tr}\left\{  \left(  D_{\mu}\mathcal{U}\right)^{\dagger}  
D^{\mu}\mathcal{U}\right\}  +
\frac{f^{2}_{\pi}}{4}\mathrm{Tr}\left\{  \chi^{\dagger}\mathcal{U} +
\mathcal{U}^{\dagger}\chi\right\}  \text{ ,} \label{app1}
\end{equation}
where $D_\mu \mathcal{U}= \partial_\mu \mathcal{U}-i \,r_\mu \,\mathcal{U}+i\,\mathcal{U}\,l_\mu$, with
external left- and right-handed vector fields $l_\mu,\, r_\mu$, respectively, and where
$\chi =2 B_0 (s+ip)$, with the LEC $B_0$ and external scalar and pseudoscalar fields $s$ and $p$, respectively.

At NLO, the number of terms increases to ten. In trace
notation, the respective Lagrangian is given by
\begin{align}
\mathscr{L}_{4}  &  = \frac{\ell_{1}}{4}\left(  \mathrm{Tr}\left\{  \left(
D_{\mu}\mathcal{U}\right)^{\dagger}D^{\mu}\mathcal{U}\right\}  \right)^{2} 
+ \frac{\ell_{2}}{4}\mathrm{Tr} \left\{\left( D_{\mu}\mathcal{U}\right)^{\dagger}  
D_{\nu}\mathcal{U}\right\}  \mathrm{Tr}\left\{  \left(D^{\mu}\mathcal{U}\right)^{\dagger}  
D^{\nu}\mathcal{U}\right\} \nonumber\\
&  \quad+\frac{h_{1} - h_{3} + \ell_{3}}{16}\left(  \mathrm{Tr}\left\{
\chi^{\dagger}\mathcal{U} + \mathcal{U}^{\dagger}\chi\right\}  \right)^{2} +
\frac{\ell_{4}}{4}\mathrm{Tr}\left\{  \left(  D_{\mu}\mathcal{U}\right)^{\dagger}
D^{\mu}\chi + \left(  D_{\mu}\chi\right)^{\dagger}  
D^{\mu}\mathcal{U}\right\} \nonumber\\
&  \quad+\ell_{5}\mathrm{Tr}\left\{  f^{(R)}_{\mu\nu}\mathcal{U}f^{(L)\mu\nu}\mathcal{U}^{\dagger}\right\} 
 - \left(  \frac{\ell_{5}}{2} + 2h_{2}\right)
\mathrm{Tr}\left\{  f^{(L)}_{\mu\nu}f^{(L)\mu\nu} + f^{(R)}_{\mu\nu}
f^{(R)\mu\nu}\right\} \nonumber\\
&  \quad+i\frac{\ell_{6}}{2}\mathrm{Tr}\left\{  f^{(R)}_{\mu\nu}\left(D^{\mu}\mathcal{U}\right)  
\left(  D^{\nu}\mathcal{U}\right)^{\dagger} +
f^{(L)}_{\mu\nu}\left(  D^{\mu}\mathcal{U}\right)^{\dagger}D^{\nu}\mathcal{U}  \right\}  
+ \frac{h_{1} - h_{3} - \ell_{7}}{16}\left(
\mathrm{Tr}\left\{  \chi^{\dagger}\mathcal{U} - \mathcal{U}^{\dagger} \chi\right\}  \right)^{2}\nonumber\\
&  \quad+\frac{h_{1} + h_{3}}{4}\mathrm{Tr}\left\{ \chi^{\dagger}\chi\right\}
- \frac{h_{1} - h_{3}}{8}\mathrm{Tr}\left\{  \chi\mathcal{U}^{\dagger}
\chi\mathcal{U}^{\dagger} + \mathcal{U}\chi^{\dagger}\mathcal{U}\chi^{\dagger}\right\}\;,  \label{app2}
\end{align}
where $f^{(L,R)}_{\mu\nu}$ are the field-strength tensors of external left- and right-handed vector fields, and
$\ell_1, \ldots, \ell_7, h_1, h_2,$ and $h_3$ are LECs (please do not confuse $h_1, h_2, h_3$ with the respective
coupling constants of the eLSM).

\subsection{Terms $\sim H\pi^{2}$ and $\sim H^{2}\pi^{2}$}

\label{appa2}

We report here the Lagrangians containing interactions of the form $H\pi^{2}$
and $H^{2}\pi^{2}$ for the heavy fields:
\begin{align}
\mathscr{L}_{\sigma_{N}\pi}  &  =\frac{1}{2}\left(  \partial_{\mu}\sigma_{N}\right)^{2}
-\frac{1}{2}m_{\sigma_{N}}^{2}\sigma_{N}^{2}-\left( \lambda_{1}+\frac{\lambda_{2}}{2}\right)  
\phi_{N}Z^{2}\sigma_{N}\vec{\pi}^{2}+g_{1}wZ^{2}\left(  \partial^{\mu}\sigma_{N}\right)  
\left(  \partial_{\mu}\vec{\pi}\right)  \cdot\vec{\pi}\nonumber\\
&  \quad+\left\{\left[g_{1}^{2}\phi _{N}+  \left(  h_{1}+h_{2}-h_{3}\right)  \frac{\phi_{N}}{2}\right]  w^{2}Z^{2}
-g_{1}wZ^{2}\right\}  \sigma_{N}\left(  \partial_{\mu}\vec{\pi}\right)^{2}
\nonumber\\
&  \quad+\left[  \frac{g_{1}^{2}}{2}+\frac{1}{4}\left(h_{1}+h_{2}-h_{3}\right)  \right]  w^{2}Z^{2}
\sigma_{N}^{2} \left(  \partial_{\mu}\vec{\pi}\right)^{2}-\frac{1}{2}\left(
\lambda_{1}+\frac{\lambda_{2}}{2}\right)  Z^{2}\sigma_{N}^{2}\vec{\pi}^{2}\text{ ,} \label{app5}
\end{align}
\begin{align}
\mathscr{L}_{a_{0}\pi}  &  =\frac{1}{2}\left(  \partial_{\mu}\vec{a}_{0}\right)^{2}
-\frac{1}{2}m_{a_{0}}^{2}\vec{a}_{0}^{2}
+\frac{\lambda_{2}}{2}Z^{2}\left(  \vec{a}_{0}\cdot\vec{\pi}\right)^{2}
-\frac{1}{2}\left(  \lambda_{1}+\frac{3\lambda_{2}}{2}\right)  Z^{2}
\vec{a}_{0}^{2}\vec{\pi}^{2}\nonumber\\
&  \quad+\frac{1}{4}\left(  h_{1}+h_{2}-h_{3}\right)
w^{2}Z^{2} \vec{a}_{0}^{2}\left(  \partial_{\mu}\vec{\pi}\right)^{2}
+\frac{g_{1}^{2}}{2}w^{2}Z^{2}\left(  \vec{a}_{0}\cdot\partial_{\mu}
\vec{\pi}\right)^{2}+\frac{h_{3}}{2}w^{2}Z^{2}\left(  \vec{a}_{0}
\times\partial_{\mu}\vec{\pi}\right)^{2}\text{ ,} \label{app6}
\end{align}
\begin{align}
\mathscr{L}_{\eta_{N}\pi}  &  =\frac{1}{2}\left(  \partial_{\mu}\eta_{N}\right)^{2}
-\frac{1}{2}m_{\eta_{N}}^{2}\eta_{N}^{2}+\left[  \frac{g_{1}^{2}}{2}
+\frac{1}{4}\left(  h_{1}+h_{2}-h_{3}\right)  \right]
w^{2}Z^{4}\left(  \partial_{\mu}\eta_{N}\right)^{2}\vec{\pi}^{2} \nonumber\\
&  \quad+\left[
\frac{g_{1}^{2}}{2}+ \frac{1}{4}\left(  h_{1}+h_{2}-h_{3}\right)  \right]
w^{2}Z^{4}\eta_{N}^{2}\left(  \partial_{\mu}\vec{\pi}\right)  ^{2}-\frac{1}{2}
\left(  \lambda_{1}+\frac{3\lambda_{2}}{2}\right)  Z^{4}\eta_{N}^{2}
\vec{\pi}^{2}\nonumber\\
&  \quad+\left(  2g_{1}^{2}+h_{2}-h_{3}\right)  w^{2}Z^{4}\eta_{N}\left(  \partial^{\mu}\eta_{N}\right)  \vec{\pi}
\cdot\left(  \partial_{\mu}\vec{\pi}\right)  +\left(  g_{3}+g_{4}\right)
w^{4}Z^{4}\left(  \partial^{\mu}\eta_{N}\right)  \left(  \partial^{\nu}
\eta_{N}\right)  \left(  \partial_{\mu}\vec{\pi}\right)  \cdot\left(
\partial_{\nu}\vec{\pi}\right) \nonumber\\
&  \quad +\left(  \frac{g_{3}}{2}+\frac{g_{4}}{2}
+\frac{g_{5}}{2}+g_{6}\right)w^{4}Z^{4}\left(  \partial_{\mu}\eta_{N}\right)^{2}
\left(  \partial_{\nu}\vec{\pi}\right)^{2}\text{ ,} \label{app7}
\end{align}
\begin{align}
\mathscr{L}_{\omega_{N}\pi}  &  =-\frac{1}{4}\omega_{N}^{\mu\nu}
\omega_{N,\mu\nu}+\frac{1}{2}m_{\omega_{N}}^{2}\omega_{N,\mu}^{2}-\frac
{\xi_{\omega_{N}}}{2}\left(  \partial_{\mu}\omega_{N}^{\mu}\right)^{2}
+\frac{1}{4}\left(  h_{1}+h_{2}+h_{3}\right)  Z^{2}\omega_{N,\mu}^{2}\vec{\pi}^{2}
 \nonumber\\
&  \quad+\left(  \frac{g_{3}}{2}+\frac{g_{4}}{2} +\frac{g_{5}}{2}+g_{6}\right)  w^{2}Z^{2}\omega_{N,\mu}^{2}
\left(  \partial_{\nu}\vec{\pi}\right)^{2}+\left(  g_{3}+g_{4}\right)
w^{2}Z^{2}\omega_{N}^{\mu}\omega_{N}^{\nu}\left(  \partial_{\mu}\vec{\pi}\right)
  \cdot\left(  \partial_{\nu}\vec{\pi}\right)  \text{ ,} \label{app8}
\end{align}
\begin{align}
\mathscr{L}_{f_{1N}\pi}  &  =-\frac{1}{4}f_{1N}^{\mu\nu}f_{1N,\mu\nu}
+\frac{1}{2}m_{f_{1N}}^{2}f_{1N,\mu}^{2}-\frac{\xi_{f_{1N}}}{2}\left(
\partial_{\mu}f_{1N}^{\mu}\right)^{2}+\left[  \frac{g_{1}^{2}}{2}+\frac{1}{4}\left(  h_{1}+h_{2}-h_{3}\right)  \right]  
Z^{2}f_{1N,\mu}^{2}\vec{\pi}^{2}\nonumber\\
&  \quad+\left(  \frac{g_{3}}{2}+\frac{g_{4}}{2}+\frac{g_{5}}{2}+g_{6}\right)
w^{2}Z^{2}f_{1N,\mu}^{2}\left(  \partial_{\nu}\vec{\pi}\right)  ^{2}+\left(
g_{3}+g_{4}\right)  w^{2}Z^{2}f_{1N}^{\mu}f_{1N}^{\nu}\left(  \partial_{\mu}\vec{\pi}\right)  
\cdot\left(  \partial_{\nu}\vec{\pi}\right)  \text{ ,}
\label{app9}
\end{align}
\begin{align}
\mathscr{L}_{\rho\pi}  &  =-\frac{1}{4}\vec{\rho}^{\,\mu\nu}
\cdot\vec{\rho}_{\mu\nu}+\frac{1}{2}m_{\rho}^{2}\vec{\rho}_{\mu}^{\,2}
-\frac{\xi_{\rho}}{2}\left(  \partial_{\mu}\vec{\rho}^{\,\mu}\right)^{2}
+g_{2}w^{2}Z^{2}\left(  \partial^{\mu}\vec{\rho}^{\,\nu}\right)
\cdot\left(  \partial_{\nu}\vec{\pi}\times\partial_{\mu}\vec{\pi}\right)
\nonumber\\
&  \quad+\left[  \left(  g_{1}^{2}\phi_{N}-h_{3}\phi_{N}\right)  wZ^{2}- g_{1}Z^{2}\right] 
 \vec{\rho}^{\,\mu}\cdot\left(  \partial_{\mu}\vec{\pi}\times\vec{\pi}\right)  
+\frac{1}{2}\left(  g_{1}^{2}-h_{3}\right)  Z^{2}\left(  \vec{\pi}\times\vec{\rho}_{\mu}\right)^{2}
+\frac{1}{4}\left(  h_{1}+h_{2}+h_{3}\right)  Z^{2}\vec{\rho}_{\mu}^{\,2}
\vec{\pi}^{2}\nonumber\\
&  \quad+\left(  -\frac{g_{3}}{2}+\frac{g_{4}}{2} +\frac{g_{5}}{2}+g_{6}\right)  w^{2}Z^{2}\vec{\rho}_{\mu}^{\,2}
\left(  \partial_{\nu}\vec{\pi}\right)  ^{2}+\left(  -g_{3}+g_{4}-g_{5}+2g_{6}\right)  
w^{2}Z^{2}\left(  \vec{\rho}^{\,\mu}\cdot\partial_{\mu}\vec{\pi}\right) 
 \left(  \vec{\rho}^{\,\nu}\cdot\partial_{\nu}\vec{\pi}\right)  \nonumber\\
&  \quad+g_{3}w^{2}Z^{2}\left[  \left(  \vec{\rho}^{\,\mu}\cdot\vec{\rho}^{\,\nu}\right)
\left(  \partial_{\mu}\vec{\pi}\right)  \cdot\left(  \partial_{\nu}\vec{\pi}\right) 
 +\left(  \vec{\rho}^{\,\mu}\cdot\partial^{\nu}\vec{\pi}\right)
\left(  \vec{\rho}_{\mu}\cdot\partial_{\nu}\vec{\pi}\right)  +\left(
\vec{\rho}^{\,\mu}\cdot\partial^{\nu}\vec{\pi}\right)  \left(  \vec{\rho}_{\nu}
\cdot\partial_{\mu}\vec{\pi}\right)  \right]  \text{ ,} \label{app10}
\end{align}
\begin{align}
\mathscr{L}_{a_{1}\pi}  &  =-\frac{1}{4}\vec{a}_{1}^{\,\mu\nu}
\cdot\vec{a}_{1,\mu\nu}+\frac{1}{2}m_{a_{1}}^{2}\vec{a}_{1,\mu}^{2}
-\frac{\xi_{a_{1}}}{2}\left(  \partial_{\mu}\vec{a}_{1}^{\,\mu}\right)^{2}
+\frac{g_{1}^{2}}{2}Z^{2}\left(  \vec{a}_{1,\mu}\cdot\vec{\pi}\right)^{2}
+\frac{1}{4}\left(  h_{1}+h_{2}-h_{3}\right)  Z^{2}\vec{a}_{1,\mu}^{2}\vec{\pi}^{2}\nonumber\\
&  \quad+\left(  -\frac{g_{3}}{2}+\frac{g_{4}}{2}+\frac{g_{5}}{2}
+g_{6}\right)  w^{2}Z^{2}\vec{a}_{1,\mu}^{2}\left(  \partial_{\nu}\vec{\pi}\right)^{2}
+\frac{h_{3}}{2}Z^{2}\left(  \vec{a}_{1,\mu}\times\vec{\pi}\right)^{2}
\nonumber\\
&  \quad+\left(  -g_{3}+g_{4}+g_{5}+2g_{6}\right)  w^{2}Z^{2}\left(  \vec{a}_{1}^{\,\mu}\cdot\partial_{\mu}\vec{\pi}\right)
\left(  \vec{a}_{1}^{\,\nu}\cdot\partial_{\nu}\vec{\pi}\right)  \nonumber\\
&  \quad+g_{3}w^{2}
Z^{2}\left[  \left(  \vec{a}_{1}^{\,\mu}\cdot\partial^{\nu}\vec{\pi}\right)
\left(  \vec{a}_{1,\mu}\cdot\partial_{\nu}\vec{\pi}\right)  
+\left(  \vec{a}_{1}^{\,\mu}\cdot\partial^{\nu}\vec{\pi}\right)  \left(  \vec{a}_{1,\nu}
\cdot\partial_{\mu}\vec{\pi}\right) + \left(  \vec{a}_{1}^{\,\mu}\cdot\vec{a}_{1}^{\,\nu}\right)
\left(  \partial_{\mu}\vec{\pi}\right)  \cdot\left(  \partial_{\nu}\vec{\pi
}\right)  \right]  \text{ .} \label{app11}
\end{align}

In the end, it is also useful to add a St\"uckelberg term for each of the
(axial-)vector mesons:
\begin{equation}
\mathscr{L}_{ST}=-\frac{\xi_{\omega_{N}}}{2}\left(  \partial_{\mu}\omega_{N}^{\mu}\right)^{2}
-\frac{\xi_{f_{1N}}}{2}\left(  \partial_{\mu}f_{1N}^{\mu}\right)^{2}-\frac{\xi_{\rho}}{2}\left(  \partial_{\mu}
\vec{\rho}^{\,\mu}\right)^{2}-\frac{\xi_{a_{1}}}{2}\left(  \partial_{\mu}\vec{a}_{1}^{\,\mu}\right)^{2}\text{ .} \label{elsm19}%
\end{equation}

\section{Other scenarios}
\label{appb}

In this appendix we consider different scenarios. We first discuss the limiting
case where vector mesons decouple (this is the case of the original LSM
with only scalar and pseudoscalar states). Then, we describe the 
case where the scalar mesons are lighter than 1 GeV [which was not favored by the fit
of Ref.\ \cite{dick}].

\subsection{Results without vector mesons}
\label{appb1}

The limit where (axial-)vector mesons decouple is realized by setting $g_{i}=0$, $i=1, \ldots,6,$
$h_i=0$, $i=1,2,3$ in the Lagrangian (\ref{elsm5}). As a consequence, $Z=1$, $w=0$,
and $\phi_{N}=f_{\pi}.$ Moreover, $c_{2,\sigma_{N}}=c_{1,\rho}=c_{2,\rho}=0$ and
\begin{equation}
c_{1,\sigma_{N}}=-\left(  \lambda_{1}+\frac{\lambda_{2}}{2}\right)  f_{\pi}\text{ .}%
\end{equation}
The corresponding low-energy coupling constants read in this limit:
\begin{align}
&  C_{1,LSM}=-\frac{1}{4}\left(  \lambda_{1}+\frac{\lambda_{2}}%
{2}\right)  +\frac{c_{1,\sigma_{N}}^{2}}{2m_{\sigma_{N}}^{2}}\left(
1-\frac{4m_{\pi}^{4}}{m_{\sigma_{N}}^{4}}\right)  \text{ ,}\\
&  C_{2,LSM}=\frac{2c_{1,\sigma_{N}}^{2}}{m_{\sigma_{N}}^{4}}\left(
1+\frac{4m_{\pi}^{2}}{m_{\sigma_{N}}^{2}}\right)  \text{ ,}\\
&  C_{3,LSM}=\frac{2c_{1,\sigma_{N}}^{2}}{m_{\sigma_{N}}^{6}}\text{ ,}\\
&  C_{4,LSM}=0\text{ .}%
\end{align}
The following comments are in order:
\begin{itemize}
\item[1)] For $m_{\sigma_N} = 1362.7$ MeV, we obtain $C_{1,LSM} = - 5.869 \pm 0.226$, which 
is more than an order of magnitude off the value in nature.
\item[2)] Upon replacing the coupling constants $\lambda_1$ and $\lambda_2$ by the masses of
the $\sigma_N$-meson and the pion, we obtain
\begin{equation}
c_{1,\sigma_{N}}^{2}=\left(  \lambda_{1}+\frac{\lambda_{2}}{2}\right)^{2}f_{\pi}^{2}
=\frac{(m_{\sigma_{N}}^{2}-m_{\pi}^{2})^{2}}{4f_{\pi}^{2}}
\end{equation}
and
\begin{equation}
C_{2,LSM}\left(  2f_{\pi}^{2}\right)  =\left( 1 - \frac{m_{\pi}^{2}}{m_{\sigma_{N}}^{2}} \right)^2
\left(  1+\frac{4m_{\pi}^{2}}{m_{\sigma_{N}}^{2}}\right)\;.
\end{equation}
Only for $m_{\sigma_{N}}\rightarrow\infty$ we recover the NLO ChPT result
\begin{equation}
C_{2,LSM}=\frac{1}{2f_{\pi}^{2}} \equiv C_{2,\chi PT}\text{ .}
\end{equation}
This result is expected from a mathematical point of view because this limit
corresponds to the nonlinear sigma model. For any finite value of
$m_{\sigma_{N}}$  the low-energy coupling constant $C_{2,LSM}$ in principle deviates from $C_{2,\chi PT}$,
but in reality, when $m_{\sigma_{N}}\gtrsim 2.3$ GeV, they still agree within errors. Including
(axial-)vector mesons, $m_{\sigma_N}$ may also be smaller.
\item[3)] $C_{3,LSM}$ receives a positive contribution from the
(pseudo)scalar sector. However, its true value is negative, cf.\ Eq.\
(\ref{c3chpt}). The (pseudo)scalar sector alone is not
sufficient to obtain agreement with data.
\item[4)] $C_{4,LSM}$ vanishes, contrary to the value obtained in nature, cf.\ Eq.\ (\ref{c4chpt}). 
This quantity depends entirely on the presence of
vector mesons. Without them, agreement with data cannot be obtained.
\end{itemize}

\subsection{Light scalars as $\bar{q}q$ states}
\label{appb2}

An interesting and important issue in the field of hadron spectroscopy is to
clarify which scalar-isoscalar state is the chiral partner of the pion. As already shown in Refs.\
\cite{denisnf2,dick,staniold,staninew}, the best fit of the parameters of the eLSM to hadron masses and decay
widths implies that the chiral partner
lies above $1\;\text{GeV}$ in mass and should be identified with the state $f_{0}(1370)$. Indeed, as
discussed by many authors [see e.g.\ Refs.\ \cite{amsler,lowscalars,tq} and the
recent review \cite{pelaezrev}], $f_{0}(500)$ should be regarded as a
four-quark state.

Yet, it is still interesting to consider the -- by now unfavored -- scenario where
$f_{0}(500)$ is (predominantly) a quark-antiquark state (and thus the chiral
partner of the pion). The scenario where the scalar states lie
below 1 GeV corresponds to the third best fit in Ref.\ \cite{dick} with a
$\chi^{2}/$d.o.f. $=11.8$. One obtains $C_{1,eLSM}= -0.4388 \pm 0.0002$ and
$C_{2,eLSM}=(9.093 \pm 0.003)\cdot10^{-5}$ MeV$^{-2}$ . 
Both values are almost a factor of two larger than in nature, but
due to the large error in Eq.\ (\ref{c1chpt}), the value of $C_{1,eLSM}$ is still acceptable. However,
that of $C_{2,eLSM}$ is further off unity than for the scenario with
the heavy scalars: $C_{2,eLSM}(2f_{\pi}^{2})\simeq 1.030672 \pm 0.000018$. Thus,
we confirm that the assignment of light scalars to ordinary quark-antiquark
mesons is less favored. This is in agreement with the recent findings of Ref.\
\cite{a0k}, in which the light scalar states $a_{0}(980)$ and $K_{0}^{\ast}(800)$ could be determined 
-- in the context of Lagrangians derived
from the eLSM -- as dynamically generated companion poles of quarkonia states
above 1 GeV in mass, namely $a_{0}(1450)$ and $K_{0}^{\ast}(1430)$.

\end{document}